\UseRawInputEncoding

\documentclass{ws-ijgmmp}

\begin{document}

\markboth{Zi-Hua Weng}
{Some field equations with composite operators in octonion spaces}

%
\catchline{}{}{}{}{}
%

\title{Some field equations with composite operators in octonion spaces
}

\author{Zi-Hua Weng
}
\address{School of Aerospace Engineering, Xiamen University, Xiamen, China
\\
College of Physical Science and Technology, Xiamen University, Xiamen, China
\\
\email{xmuwzh@xmu.edu.cn
}
}


\maketitle

\begin{history}
\received{(Day Month Year)}
\revised{(Day Month Year)}
\end{history}

\begin{abstract}
The paper aims to explore the impact of composite operators containing a few physical quantities on the gravitational and electromagnetic fields, studying the influencing factors and physical properties of octonion field equations. J. C. Maxwell first utilized the quaternions and vector terminology to describe the electromagnetic fields. The octonions can be used to simultaneously describe the physical quantities of electromagnetic and gravitational fields, including the octonion field potential, field strength, field source, linear momentum, angular momentum, torque and force. In the octonion spaces, the field strength and quaternion operator are able to combine together to become one composite operator, making an important contribution to the field equations. Similarly, the field potential can also form some composite operators with the quaternion operator, and they have a certain impact on the field equations. Furthermore, other physical quantities can also be combined with the quaternion operator to form several composite operators. In these composite operators, multiple physical quantities can also have a certain impact on the field equations. In other words, the field strength does not occupy a unique central position, compared to other physical quantities in these composite operators. In the field theories described by the octonions, various field equations can be derived from the application of different composite operators. According to different composite operators, it is able to infer the field equations when the field strength or/and field potential make a certain contribution, and it is also possible to deduce several new field equations when the remaining physical quantities play a certain role. This further deepens the understanding of the physical properties of field equations.
\end{abstract}


\section{\label{sec:level1}Introduction}

Can the field potential exert an impact on the electromagnetic and gravitational equations? May multiple physical quantities have an effect on the composite operators of field equations? Are there other types of field equations? For a long time, these difficult issues have puzzled and attracted the scholars. It was not until the emergence of the octonion field theory (short for field theory described by the octonions) that these questions were partially answered. In the octonion field theory, the field potential and other physical quantities are able to contribute to the electromagnetic and gravitational equations. The Maxwell equations, which are influenced by the field strength, are only one of some special cases of the octonion field equations. Further there exist several field equations that are influenced by other types of physical quantities.

When establishing the electromagnetic theory, J. C. Maxwell believed that the electromagnetic potential is a fundamental quantity to describe the electromagnetic fields. Later, H. R. Hertz and O. Heaviside and others considered that the electromagnetic strength is the fundamental quantity of electromagnetic fields, while the electromagnetic potential is merely an auxiliary quantity. It is the viewpoint that has been inherited so far in the classical electrodynamics. This viewpoint is positive for H. R. Hertz and O. Heaviside and others. According to this point of view, the electromagnetic equations, relevant to the algebra of quaternions, can be rewritten into the Maxwell equations described by the vector terminology. Obviously, the field strength plays an important role in the electromagnetic fields, the point of view can be extended from the electromagnetic fields to the gravitational fields.

However, Maxwell's viewpoint received renewed attention, in modern times. It breeds the new content, which is the gauge field. In 1918, H. Weyl introduced the concept of gauge transformation. In 1954, C. N. Yang and R. L. Mills proposed the gauge field theory. Both the electromagnetic theory and gauge field theory possess the locally invariant. However, there is no direct interaction between two photons, while there is a direct interaction between two quanta of gauge fields.

In 1967, S. L. Glashow utilized the group, SU(2)×U(1), to describe the weak and electromagnetic interactions. S. Weinberg and A. Salam applied the Higgs mechanism to the gauge field theory. In the electroweak unified theory, many predictions have been tested experimentally. Apparently, the field potential plays an important role in the gauge fields, and the point of view can be extended into the electromagnetic and gravitational fields in the paper.

The classical theories of electromagnetic and gravitational fields have achieved many achievements, including the quantum properties of microscopic particles in the electromagnetic fields. Nevertheless, there are some difficult problems in the classical theory of electromagnetic and gravitational fields.

(1) Redundant field potential. The classical field theories deem that the field potential is purely a mathematical tool and has no physical significance. In the classical electromagnetic and gravitational theories, the field strength is a fundamental and observable physical quantity. Nonetheless, the field potential is just a dispensable thing introduced for the convenience of calculation. The field potential has only mathematical significance and does not represent any physical essence.

(2) Remaining physical quantities. The classical field theories assume that the field strength will play an important role. But the field potential is unable to exert any influence, in the classical electromagnetic and gravitational fields. Further, other physical quantities, except for the field potential and field strength, are incapable of combining with the Hamiltonian operator to form the composite operators. In the composite operators, these remaining physical quantities may not possess any physical impact.

(3) Classical field equations. In the classical electromagnetic and gravitational theories, there exists only a composite operator composed of the Hamiltonian operator and field strength. But there is no composite operator composed of the Hamiltonian operator and field potential. Either there is no composite operator composed of the Hamiltonian operator and other physical quantities. Consequently, it is impossible to deduce a new set of field equations when the remaining physical quantities have a certain impact.

The above shows that the existing research on the exploration of classical field theories has some shortcomings in the electromagnetic and gravitational media. This defect limits the applicability of some research results related to the field equations in the classical field theories. In contrast, the application of octonions is able to solve several puzzles related with the material media, explaining some difficult problems derived from the classical field theories.

W. R. Hamilton invented the algebra of quaternions in 1843. The quaternions possess one real basis vector and three imaginary basis vectors \cite{flaut}. Afterwards, J. T. Graves and A. Cayley discovered independently the octonions. It is called as the classical octonion. The octonions own one real basis vector and seven imaginary basis vectors. At that time, the engineers believed that the quaternions were too complicated and inconvenient to use, they decomposed the quaternions into the scalars and vectors. In 1876, J. C. Maxwell utilized simultaneously the quaternions and vector terminology to study the physical properties of electromagnetic fields. This method has been passed down to this day. The contemporary scholars apply the quaternions and octonions to explore the electromagnetic and gravitational theories. The paper utilizes the classical octonions to research the physical properties of electromagnetic and gravitational fields.

J. C. Maxwell first studied the physical properties of electromagnetic fields by means of the quaternions. The contemporary scholars \cite{gogberashvili} apply the algebra of quaternions \cite{goldman,rawat} and octonions \cite{mironov,deleo} to explore the electromagnetic fields \cite{tanisli1}, gravitational fields, relativity \cite{moffat}, quantum mechanics \cite{morita,bernevig}, dark matter \cite{furui}, strong nuclear fields \cite{chanyal2,furey}, weak nuclear fields \cite{majid,farrill}, black holes \cite{bossard}, fluids \cite{tanisli2,kansu} and plasmas \cite{demir} and others.

R. Descartes considered that the space is the extension of substance. Nowadays, the Cartesian thought can be improved to that the fundamental space is the extension of fundamental field. In the paper, the fundamental fields consist of the gravitational field and electromagnetic field. Each fundamental field possesses one fundamental space. It means that the fundamental space for the electromagnetic fields is independent of that for gravitational fields. Further, each fundamental space is one of quaternion spaces. It is able to choose an octonion space, to contain two independent quaternion spaces. In other words, the application of octonion spaces is able to explore simultaneously the physical properties of electromagnetic and gravitational fields within material media. Moreover, there are other methods to be able to introduce the octonions into the field theories, exploring the physical properties of electromagnetic and gravitational fields.

Through comparison and analysis, it is able to find some significant characteristics of the contribution of various physical quantities in the octonion field theory.

(1) Practical field potential. The octonion field theory holds that the field potential has physical significance. In the octonion field theory, the field strength is a fundamental and observable physical quantity. Meanwhile, the field potential is an indispensable physical quantity, which facilitates the theoretical calculations. The field potential is able to simplify mathematical operations, and is the physical quantity with observational effects.

(2) Several composite operators. In the octonion field theory, not only the field strength but also the field potential can play a certain role. Further, the remaining physical quantities and quaternion operator are able to combine together to become some composite operators. In these composite operators, the remaining physical quantities can also exert a certain impact. For different physical quantities, their associated composite operators may be different from each other.

(3) New field equations. There are various composite operators in the octonion field theory. Some different field equations can be derived from these composite operators. By means of different composite operators, it is possible to infer some field equations when the field strength and field potential play a certain role. Meanwhile a few new field equations will be deduced in case the remaining physical quantities have a certain impact.

In the paper, the octonion field strength and quaternion operator are able to combine together to become a composite operator, exploring simultaneously the physical properties of electromagnetic and gravitational fields. The quaternion operator and octonion field potential can form some composite operators, studying the contribution of octonion field potentials to the octonion field equations. Meanwhile, the octonion field strength, octonion field potential, and quaternion operator can constitute some composite operators, researching the contribution of field strength and field potential to the octonion field equations. Furthermore, other physical quantities and quaternion operator can form some composite operator in the octonion spaces, investigating the contribution of other physical quantities to the octonion field equations.

\begin{table}[h]
\caption{The multiplication table of octonion.}
\label{tab:table3}
\centering
\begin{tabular}{ccccccccc}
\hline\noalign{\smallskip}
$ $                   & $1$                   & $\emph{\textbf{i}}_1$  & $\emph{\textbf{i}}_2$   & $\emph{\textbf{i}}_3$   & $\emph{\textbf{I}}_0$  & $\emph{\textbf{I}}_1$  & $\emph{\textbf{I}}_2$  & $\emph{\textbf{I}}_3$  \\
\noalign{\smallskip}\hline\noalign{\smallskip}
$1$                   & $1$                   & $\emph{\textbf{i}}_1$  & $\emph{\textbf{i}}_2$   & $\emph{\textbf{i}}_3$   & $\emph{\textbf{I}}_0$  & $\emph{\textbf{I}}_1$  & $\emph{\textbf{I}}_2$  & $\emph{\textbf{I}}_3$  \\
$\emph{\textbf{i}}_1$ & $\emph{\textbf{i}}_1$ & $-1$                   & $\emph{\textbf{i}}_3$   & $-\emph{\textbf{i}}_2$  & $\emph{\textbf{I}}_1$  & $-\emph{\textbf{I}}_0$ & $-\emph{\textbf{I}}_3$ & $\emph{\textbf{I}}_2$  \\
$\emph{\textbf{i}}_2$ & $\emph{\textbf{i}}_2$ & $-\emph{\textbf{i}}_3$ & $-1$                    & $\emph{\textbf{i}}_1$   & $\emph{\textbf{I}}_2$  & $\emph{\textbf{I}}_3$  & $-\emph{\textbf{I}}_0$ & $-\emph{\textbf{I}}_1$ \\
$\emph{\textbf{i}}_3$ & $\emph{\textbf{i}}_3$ & $\emph{\textbf{i}}_2$  & $-\emph{\textbf{i}}_1$  & $-1$                    & $\emph{\textbf{I}}_3$  & $-\emph{\textbf{I}}_2$ & $\emph{\textbf{I}}_1$  & $-\emph{\textbf{I}}_0$ \\
$\emph{\textbf{I}}_0$ & $\emph{\textbf{I}}_0$ & $-\emph{\textbf{I}}_1$ & $-\emph{\textbf{I}}_2$  & $-\emph{\textbf{I}}_3$  & $-1$                   & $\emph{\textbf{i}}_1$  & $\emph{\textbf{i}}_2$  & $\emph{\textbf{i}}_3$  \\
$\emph{\textbf{I}}_1$ & $\emph{\textbf{I}}_1$ & $\emph{\textbf{I}}_0$  & $-\emph{\textbf{I}}_3$  & $\emph{\textbf{I}}_2$   & $-\emph{\textbf{i}}_1$ & $-1$                   & $-\emph{\textbf{i}}_3$ & $\emph{\textbf{i}}_2$  \\
$\emph{\textbf{I}}_2$ & $\emph{\textbf{I}}_2$ & $\emph{\textbf{I}}_3$  & $\emph{\textbf{I}}_0$   & $-\emph{\textbf{I}}_1$  & $-\emph{\textbf{i}}_2$ & $\emph{\textbf{i}}_3$  & $-1$                   & $-\emph{\textbf{i}}_1$ \\
$\emph{\textbf{I}}_3$ & $\emph{\textbf{I}}_3$ & $-\emph{\textbf{I}}_2$ & $\emph{\textbf{I}}_1$   & $\emph{\textbf{I}}_0$   & $-\emph{\textbf{i}}_3$ & $-\emph{\textbf{i}}_2$ & $\emph{\textbf{i}}_1$  & $-1$                   \\
\noalign{\smallskip}\hline
\end{tabular}
\end{table}

\section{\label{sec:level1}Octonion properties}

The octonion $\mathbb{O}$ is a non-associative extension of the quaternion $\mathbb{H}$ , belonging to the non-associative algebra. The octonions were first described in 1844, within a letter from J. T. Graves to W. R. Hamilton. Subsequently, the octonions were independently published by A. Cayley in 1845. Every octonion is a linear combination of unit octonions. The addition of octonions involves adding the corresponding coefficients, just like the addition of complex numbers or of quaternions. According to the linear properties, the product of octonions is entirely determined by the multiplication table of unit octonions (Table 1).

A more systematic method for defining octonions is through the Cayley-Dickson construction of dimension 2n. Just as the quaternion can be defined by a pair of complex numbers, the octonion can be defined by a pair of quaternions.

The product of two pairs of quaternions $(\mathbb{C}_1, \mathbb{C}_2)$ and $(\mathbb{C}_3, \mathbb{C}_4)$ is defined as,
\begin{eqnarray}
(\mathbb{C}_1 , \mathbb{C}_2) \circ (\mathbb{C}_3 , \mathbb{C}_4) = (\mathbb{C}_1 \mathbb{C}_3 - \mathbb{C}_4^\ast \mathbb{C}_2 , \mathbb{C}_4 \mathbb{C}_1 + \mathbb{C}_2 \mathbb{C}_3^\ast) ~,
\end{eqnarray}
where $\mathbb{C}_j$ is a quaternion. $\mathbb{C}_j^\ast$ is the conjugate quaternion. $\circ$ denotes the octonion multiplication. $j = 0, 1, 2, 3$.

In terms of an octonion $\mathbb{O}$ , the basis vector is $( \textbf{i}_j , \textbf{I}_j )$, the coefficients are $z_j$ and $Z_j$ . The octonion $\mathbb{Z}$ can be written as,
\begin{eqnarray}
\mathbb{Z} = \Sigma ( \textbf{i}_j z_j + \textbf{I}_j Z_j ) ~,
\end{eqnarray}
where $\textbf{i}_0 = 1$. $\textbf{i}_k^2 = -1$. $\textbf{I}_j^2 = -1$. $k = 1, 2, 3$.

The conjugate octonion $\mathbb{Z}^*$ can be defined as,
\begin{eqnarray}
\mathbb{Z}^* = z_0 + \Sigma ( \textbf{i}_k^\ast z_k + \textbf{I}_j^\ast Z_j ) ~,
\end{eqnarray}
where $\ast$ denotes the octonion conjugate. $\textbf{i}_k^\ast = - \textbf{i}_k$ . $\textbf{I}_j^\ast = - \textbf{I}_j$ .

The norm $Z^2$ of octonion $\mathbb{Z}$ can be defined as,
\begin{eqnarray}
Z^2 = \mathbb{Z} \circ \mathbb{Z}^\ast = \Sigma ( z_j^2 + Z_j^2 ) ~,
\end{eqnarray}
where the norm is consistent with the standard Euclidean norm in the real number field.

In 1898, A. Hurwitz proved a theorem, that is, Hurwitz's theorem. The theorem claims that any normed division algebra with identity elements is isomorphic to one of four algebras: R, $\textbf{C}$, $\mathbb{H}$, and $\mathbb{O}$ , they represent the real numbers, complex numbers, quaternions, and octonions, respectively. In other words, the only normed division algebras are R, $\textbf{C}$, $\mathbb{H}$, and $\mathbb{O}$, on the real number field. These four algebras also form the only alternating and finite-dimensional division algebra in the real number field. Since the octonions are not associative, the non-zero elements of octonion $\mathbb{O}$ do not form a group.

According to the multiplication table of octonions, the multiplication of octonions is neither commutative nor associative. However, the octonions do satisfy a weaker form of associative property, which is the alternating property. It means that the subalgebra generated by any two elements is associative. In fact, the subalgebra generated by any two elements of octonion $\mathbb{O}$ is isomorphic to R, $\textbf{C}$, or $\mathbb{H}$, and they are all associative.

The algebra of octonions is able to explore the physical properties of gravitational and electromagnetic fields.

\section{\label{sec:level1}Field strength}

The classical octonions can be applied to simultaneously research the physical quantities of electromagnetic and gravitational fields within material media, including the octonion field source, torque, and force and others.

The octonion space can be applied to describe the physical properties of gravitational and electromagnetic fields simultaneously. In the octonion space $\mathbb{O}$ , the basis vector is $( \textbf{i}_j , \textbf{I}_j )$ , the coordinate value is $( i r_0 , r_k , i R_0 , R_k )$ . The octonion radius vector is, $\mathbb{R} = i r_0 \textbf{i}_0 + \Sigma r_k \textbf{i}_k + k_{eg} ( i R_0 \textbf{I}_0 + \Sigma R_k \textbf{I}_k )$ , the octonion velocity is, $\mathbb{V} = i v_0 \textbf{i}_0 + \Sigma v_k \textbf{i}_k + k_{eg} ( i V_0 \textbf{I}_0 + \Sigma V_k \textbf{I}_k )$. Herein $k_{eg}$ is a coefficient to meet the requirements of dimensional homogeneity. $r_0 = v_0 t$ . $v_0$ is the speed of light, and $t$ is the time. $r_j$ , $R_j$ , $v_j$ , and $V_j$ are all real.

In the octonion spaces, the quaternion operator $\lozenge$ and octonion field strength $\mathbb{F}$ can be combined together to become one composite operator, $\lozenge_B = ( i \mathbb{F} / v_0 + \lozenge )$. The composite operator is able to act on the physical quantities, achieving some octonion physical quantities related to the contribution of octonion field strength \cite{weng3}. By means of two operators, $\lozenge$ and $\lozenge_B$ , it is capable of defining the octonion field source $\mathbb{S}$ , linear momentum $\mathbb{P}$ , angular momentum $\mathbb{L}$ , torque $\mathbb{W}$ , and force $\mathbb{N}$ (Table 2). Herein $\lozenge = i \textbf{i}_0 \partial_0 + \Sigma \textbf{i}_k \partial_k$ . $\nabla = \Sigma \textbf{i}_k \partial_k$ . $\partial_j = \partial / \partial r_j$ .

In the octonion spaces, the octonion field source $\mathbb{S}$ can be defined as \cite{weng1}
\begin{eqnarray}
\mu \mathbb{S} = - ( i \mathbb{F} / v_0 + \lozenge )^\ast \circ \mathbb{F} ~,
\end{eqnarray}
where the octonion field strength is, $\mathbb{F} = \lozenge \circ \mathbb{A}$ , with $\mathbb{A}$ being the octonion field potential. $\mu$ is the coefficient.

The octonion torque $\mathbb{W}$ is defined as,
\begin{eqnarray}
\mathbb{W} = - v_0 ( i \mathbb{F} / v_0 + \lozenge ) \circ  \mathbb{L}    ~,
\end{eqnarray}
where the octonion angular momentum is, $\mathbb{L} = ( \mathbb{R} + k_{rx} \mathbb{X} )^\times \circ \mathbb{P}$ . $\mathbb{X}$ is the octonion integrating function of field potential. The octonion linear momentum is, $\mathbb{P} = \mu \mathbb{S} / \mu_g$ . $k_{rx}$ is a coefficient, to meet the requirements of dimensional homogeneity physically. $\times$ is the complex conjugate. $\mu_g$ is the gravitational constant.

The octonion force $\mathbb{N}$ is defined as,
\begin{eqnarray}
\mathbb{N} = - ( i \mathbb{F} / v_0 + \lozenge ) \circ \mathbb{W} ~.
\end{eqnarray}

In the octonion spaces, when the above meets certain conditions, it is able to achieve some continuity equations and equilibrium equations \cite{weng2} , includes the fluid continuity equation, current continuity equation, and force equilibrium equation and so forth. It weaves together some laws and theorems, that were originally isolated from each other, for mutual verification or correction.

The above method can be extended from the case of octonion field strength to that of octonion field potential.

\begin{table}[h]
\centering
\caption{The quaternion operator and octonion field strength can be combined together to become one composite operator, achieving some octonion physical quantities related to the contribution of octonion field strength.}
\begin{tabular}{@{}ll@{}}
\hline\noalign{\smallskip}
octonion physical quantity~~~~~~~~             &   definition                                                                                                                \\
\noalign{\smallskip}\hline\noalign{\smallskip}
octonion integrating function                  &   $\mathbb{X} = \mathbb{X}_g + k_{eg} \mathbb{X}_e$                                                                         \\
octonion field potential                       &   $\mathbb{A} = i \lozenge^\times \circ \mathbb{X}$                                                                         \\
octonion field strength                        &   $\mathbb{F} = \lozenge \circ \mathbb{A}$                                                                                  \\
octonion field source                          &   $\mu \mathbb{S} = - ( i \mathbb{F} / v_0 + \lozenge )^\ast \circ \mathbb{F}$                                              \\
octonion linear momentum                       &   $\mathbb{P} = \mu \mathbb{S} / \mu_g$                                                                                     \\
octonion angular momentum                      &   $\mathbb{L} = ( \mathbb{R} + k_{rx} \mathbb{X} )^\times \circ \mathbb{P}$                                                 \\
octonion torque                                &   $\mathbb{W} = - v_0 ( i \mathbb{F} / v_0 + \lozenge ) \circ  \mathbb{L}  $        \\
octonion force                                 &   $\mathbb{N} = - ( i \mathbb{F} / v_0 + \lozenge ) \circ \mathbb{W}$                                                       \\
\noalign{\smallskip}\hline
\end{tabular}
\end{table}

\section{\label{sec:level1}Field potential}

Y. Aharonov and D. Bohm proposed the Aharonov-Bohm effect (A-B effect for short ) in 1959. A-B effect emphasizes the importance of electromagnetic potential, revealing the contribution of electromagnetic potential to the phase of electron wave function. There is one special type of situation. In the particular space through which the moving electrons pass, there is only the electromagnetic potential but no electromagnetic strength. The experiment cleverly proves that the electromagnetic potential possesses the physical significance \cite{a-b}. The scholars have constructed two experimental schemes, including the electric A-B effect and magnetic A-B effect. The magnetic A-B effect has been achieved, while the electrical A-B effect has not been finished.

Y. Aharonov and D. Bohm not only conducted some theoretical calculations, but also designed a few explicit validation experiments. Their ideas and thought experiment have attracted wide attention. Some scholars have completed a few experiments related to this, and achieved the expected results. However, several scholars always argue about these results. They believe that the theory is flawed, and there may also be magnetic leakage in the experiments. In 1986, A. Tonomura et al. accomplished the rigorous experiments without magnetic leakage \cite{tonomura}, resulting in the validation of the A-B effect in the academic community. The A-B effect was also verified in some experiments related to superconductors and others \cite{berry}. Furthermore, M. A. Hohensee et al. in 2012 conceived an experimental plan to measure the A-B effect in the gravitational fields \cite{hohensee}. C. Overstreet et al. detected the A-B effect within the gravitational fields in 2022 \cite{overstreet}.

The gauge field theory and electroweak unified theory reveal that the Hamiltonian operator and field potential can form a composite operator. This inspires us that the octonion field potential and quaternion operator will be combined together to become some composite operators. It is similar to that the octonion field strength and quaternion operator may constitute a composite operator.

In the octonion field theory, the octonion field potential can be utilized to replace the octonion field strength in some composite operators. The quaternion operator and octonion field potential can comprise several composite operators. These composite operators can act on multiple physical quantities, obtaining some octonion physical quantities related to the contribution of the octonion field potential (Table 3).

The quaternion operator $\lozenge$ and octonion field potential $\mathbb{A}$ can constitute a few composite operators. By means of these composite operators, it is able to define separately several octonion physical quantities related to the contribution of the octonion field potential, including the octonion field strength $\mathbb{F}$ , field source $\mathbb{S}$ , torque $\mathbb{W}$ , and force $\mathbb{N}$ .

In the octonion spaces, the octonion field strength $\mathbb{F}$ is defined as,
\begin{eqnarray}
\mathbb{F} = ( \lozenge +  k_{aa} \mathbb{A} ) \circ \mathbb{A} ~,
\end{eqnarray}
where $k_{aa}$ is one coefficient, to meet the requirements of dimensional homogeneity.

The octonion field source $\mathbb{S}$ can be defined as,
\begin{eqnarray}
\mu \mathbb{S} = - ( \lozenge + k_{fa} \mathbb{A} )^\ast \circ \mathbb{F} ~,
\end{eqnarray}
where $k_{fa}$ is one coefficient, to satisfy the needs of dimensional homogeneity.

The octonion torque $\mathbb{W}$ can be defined as,
\begin{eqnarray}
\mathbb{W} = - v_0 ( \lozenge  + k_{la} \mathbb{A} ) \circ  \mathbb{L}  ~,
\end{eqnarray}
where $k_{la}$ is one coefficient, to meet the requirements of dimensional homogeneity. $\mathbb{L} = ( \mathbb{R} + k_{rx} \mathbb{X} )^\times \circ \mathbb{P}$ . $\mathbb{P} = \mu \mathbb{S} / \mu_g$ .

The octonion force $\mathbb{N}$ can be defined as,
\begin{eqnarray}
\mathbb{N} = - ( \lozenge + k_{wa} \mathbb{A} ) \circ \mathbb{W} ~,
\end{eqnarray}
where $k_{wa}$ is one coefficient, to satisfy the needs of dimensional homogeneity.

In the octonion spaces, these equations can be expanded further when certain conditions are met in the above. The expansion methods of Eqs.(8) to (11) are similar to these of Eqs.(5) to (7) (see Ref. [21]). For instance, we may expand Eq.(9) into eight equations independent of each other \cite{weng4} , by replacing the composite operator $( i \mathbb{F} / v_0 + \lozenge )$ with $( \lozenge + k_{fa} \mathbb{A} )$. And the expansion method of Eq.(9) is similar to that of Eq.(5).

The above method can be extended from the case of the octonion field potential into that the octonion field potential and field strength jointly affect other physical quantities.

\begin{table}[h]
\centering
\caption{The quaternion operator and octonion field potential can be combined together to become a few composite operators, achieving several octonion physical quantities related to the contribution of octonion field potential.}
\begin{tabular}{@{}ll@{}}
\hline\noalign{\smallskip}
octonion physical quantity~~~~~~~              &   definition                                                                                                                 \\
\noalign{\smallskip}\hline\noalign{\smallskip}
octonion field strength                        &   $\mathbb{F} = ( \lozenge +  k_{aa} \mathbb{A} ) \circ \mathbb{A}$                                                          \\
octonion field source                          &   $\mu \mathbb{S} = - ( \lozenge + k_{fa} \mathbb{A} )^\ast \circ \mathbb{F}$                                                \\
octonion torque                                &   $\mathbb{W} = - v_0 ( \lozenge  + k_{la} \mathbb{A} ) \circ  \mathbb{L} $          \\
octonion force                                 &   $\mathbb{N} = - ( \lozenge + k_{wa} \mathbb{A} ) \circ \mathbb{W}$                                                         \\
\noalign{\smallskip}\hline
\end{tabular}
\end{table}

\section{Field equations of field potentials}

In the octonion spaces, the octonion field potential is, $\mathbb{A} = \mathbb{A}_g + k_{eg} \mathbb{A}_e$ . When the octonion field potential $\mathbb{A}$ plays an important role, the octonion field strength, $\mathbb{F} = ( \lozenge +  k_{aa} \mathbb{A} ) \circ \mathbb{A}$ , can be written as,
\begin{eqnarray}
\mathbb{F} = \mathbb{F}_g + k_{eg} \mathbb{F}_e  ~,
\end{eqnarray}
where the gravitational potential is $\mathbb{A}_g$ with the basis vector $\textbf{i}_j$ , and the electromagnetic potential is $\mathbb{A}_e$ with the basis vector $\textbf{I}_j$ . $\mathbb{F}_g = \lozenge \circ \mathbb{A}_g + k_{aa} ( \mathbb{A}_g \circ \mathbb{A}_g + k_{eg}^2 \mathbb{A}_e \circ \mathbb{A}_e )$, while $\mathbb{F}_e = \lozenge \circ \mathbb{A}_e + k_{aa} k_{eg} ( \mathbb{A}_g \circ \mathbb{A}_e + \mathbb{A}_e \circ \mathbb{A}_g )$ .

The octonion field source $\mathbb{S}$ of the electromagnetic and gravitational fields is written as,
\begin{eqnarray}
\mu \mathbb{S} && = - ( \lozenge +  k_{fa} \mathbb{A} )^\ast \circ \mathbb{F}
\nonumber
\\
&& = \mu_g \mathbb{S}_g + k_{eg} \mu_e \mathbb{S}_e - k_{fa} \mathbb{A}^\ast \circ \mathbb{F} ~,
\end{eqnarray}
where $\mathbb{A}^\ast \circ \mathbb{F} = ( \mathbb{A}_g^\ast \circ \mathbb{F}_g + k_{eg}^2 \mathbb{A}_e^\ast \circ \mathbb{F}_e ) + k_{eg} ( \mathbb{A}_g^\ast \circ \mathbb{F}_e + \mathbb{A}_e^\ast \circ \mathbb{F}_g )$ . $\mu_g \mathbb{S}_g = - \lozenge^\ast \circ \mathbb{F}_g $ , and $\mu_e \mathbb{S}_e = - \lozenge^\ast \circ \mathbb{F}_e $ . $\mu_e$ is the electromagnetic constant.

The octonion linear momentum, $\mathbb{P} = \mu \mathbb{S} / \mu_g$ , is written as,
\begin{eqnarray}
\mathbb{P} = \mathbb{P}_g + k_{eg} \mathbb{P}_e ~,
\end{eqnarray}
where
$\mathbb{P}_g = \{ \mu_g \mathbb{S}_g - k_{fa} ( \mathbb{A}_g^\ast \circ \mathbb{F}_g + k_{eg}^2 \mathbb{A}_e^\ast \circ \mathbb{F}_e ) \} / \mu_g$ , $\mathbb{P}_e = \{ \mu_e \mathbb{S}_e - k_{fa} ( \mathbb{A}_g^\ast \circ \mathbb{F}_e + \mathbb{A}_e^\ast \circ \mathbb{F}_g ) \} / \mu_g$ .

In the octonion spaces, the composite radius vector is $\mathbb{R}^+ = \mathbb{R} + k_{rx} \mathbb{X}$ , with $\mathbb{A} = i \lozenge^\times \circ \mathbb{X}$ . The octonion angular momentum, $\mathbb{L} = (\mathbb{R}^+)^\times \circ \mathbb{P}$ , is defined as
\begin{eqnarray}
\mathbb{L} = \mathbb{L}_g + k_{eg}^\times \mathbb{L}_e ~,
\end{eqnarray}
where $\mathbb{L}_g = (\mathbb{R}_g^+)^\times \circ \mathbb{P}_g + k_{eg}^2 (\mathbb{R}_e^+)^\times \circ \mathbb{P}_e$ , and $\mathbb{L}_e = (\mathbb{R}_g^+)^\times \circ \mathbb{P}_e + (\mathbb{R}_e^+)^\times \circ \mathbb{P}_g$ . $\mathbb{R}_g^+ = \mathbb{R}_g + k_{rx} \mathbb{X}_g$ . $\mathbb{R}_e^+ = \mathbb{R}_e + k_{rx} \mathbb{X}_e$ . $\mathbb{X} = \mathbb{X}_g + k_{eg} \mathbb{X}_e$ .

From the octonion angular momentum $\mathbb{L}$, the octonion torque, $\mathbb{W} = - v_0 ( \lozenge + k_{la} \mathbb{A}  ) \circ  \mathbb{L} $ , can be written as
\begin{equation}
\mathbb{W} = \mathbb{W}_g + k_{eg} \mathbb{W}_e ~,
\end{equation}
where $\mathbb{W}_g = - v_0 \{ \lozenge  \circ \mathbb{L}_g + k_{la} ( \mathbb{A}_g \circ \mathbb{L}_g + k_{eg}^2 \mathbb{A}_e \circ \mathbb{L}_e ) \} $ , $\mathbb{W}_e = - v_0 \{ \lozenge  \circ \mathbb{L}_e + k_{la} ( \mathbb{A}_g \circ \mathbb{L}_e + \mathbb{A}_e \circ \mathbb{L}_g ) \} $ .

The octonion force, $\mathbb{N} = - ( \lozenge + k_{wa} \mathbb{A}  ) \circ \mathbb{W}$ , can be written as
\begin{equation}
\mathbb{N} = \mathbb{N}_g + k_{eg} \mathbb{N}_e ~,
\end{equation}
where $\mathbb{N}_g = - \{ \lozenge  \circ \mathbb{W}_g + k_{wa} ( \mathbb{A}_g \circ \mathbb{W}_g + k_{eg}^2 \mathbb{A}_e \circ \mathbb{W}_e ) \} $ , $\mathbb{N}_e = - \{ \lozenge  \circ \mathbb{W}_e + k_{wa} ( \mathbb{A}_g \circ \mathbb{W}_e + \mathbb{A}_e \circ \mathbb{W}_g ) \} $ .

The above states that the field equations, when the field potential plays an crucial role, will be distinct from these when the field strength acts as an important part, in the electromagnetic and gravitational fields.

\section{\label{sec:level1}Field strength and field potential}

In the octonion spaces, the quaternion operator $\lozenge$ , octonion field potential $\mathbb{A}$ , and octonion field strength $\mathbb{F}$ can be combined together to become some composite operators. These operators can act on a few physical quantities, achieving several octonion physical quantities related to the contribution of octonion field potential and field strength simultaneously (Table 4), including the octonion field source $\mathbb{S}$ , octonion torque $\mathbb{W}$ , and octonion force $\mathbb{N}$ .

In the octonion spaces, the octonion field source $\mathbb{S}$ will be defined as,
\begin{eqnarray}
\mu \mathbb{S} = - ( \lozenge + k_{fa} \mathbb{A} + k_{ff} \mathbb{F} )^\ast \circ \mathbb{F} ~,
\end{eqnarray}
where $k_{ff}$ is one coefficient, to satisfy the needs of dimensional homogeneity.

The octonion torque $\mathbb{W}$ will be defined as,
\begin{eqnarray}
\mathbb{W} = - v_0 ( \lozenge + k_{la} \mathbb{A} + k_{lf} \mathbb{F} ) \circ  \mathbb{L}  ~,
\end{eqnarray}
where $k_{lf}$ is one coefficient, to meet the requirements of dimensional homogeneity.

The octonion force $\mathbb{N}$ will be defined as,
\begin{eqnarray}
\mathbb{N} = - ( \lozenge + k_{wa} \mathbb{A} + k_{wf} \mathbb{F} ) \circ \mathbb{W} ~,
\end{eqnarray}
where $k_{wf}$ is one coefficient, to satisfy the needs of dimensional homogeneity.

In the octonion spaces, these above equations can be expanded further when certain conditions are met. The expansion methods of Eqs.(18) to (20) are similar to these of Eqs.(5) to (7). For example, one can expand Eq.(19) into eight equations independent of each other, by replacing the composite operator $( i \mathbb{F} / v_0 + \lozenge )$ with $( \lozenge + k_{la} \mathbb{A} + k_{lf} \mathbb{F} )$. In particular, the expansion method of Eq.(19) is similar to that of Eq.(6).

The above method can be extended from the case of octonion field potential and field strength to the case where multiple octonion physical quantities collectively exert their influence.

\begin{table}[h]
\centering
\caption{The quaternion operator, octonion field potential, and octonion field strength can be combined together to become a few composite operators, achieving several octonion physical quantities related to the contribution of octonion field potential and field strength simultaneously.}
\begin{tabular}{@{}ll@{}}
\hline\noalign{\smallskip}
octonion physical quantity~~~~~~~              &   definition                                                                                                                                         \\
\noalign{\smallskip}\hline\noalign{\smallskip}
octonion field source                          &   $\mu \mathbb{S} = - ( \lozenge + k_{fa} \mathbb{A} + k_{ff} \mathbb{F} )^\ast \circ \mathbb{F}$                                                    \\
octonion torque                                &   $\mathbb{W} = - v_0 ( \lozenge + k_{la} \mathbb{A} + k_{lf} \mathbb{F} ) \circ  \mathbb{L} $               \\
octonion force                                 &   $\mathbb{N} = - ( \lozenge + k_{wa} \mathbb{A} + k_{wf} \mathbb{F} ) \circ \mathbb{W}$                                                             \\
\noalign{\smallskip}\hline
\end{tabular}
\end{table}

\section{\label{sec:level1}Multiple physical quantities}

In the octonion spaces, the quaternion operator, along with the octonion field potential and octonion field strength, can constitute some composite operators. Further the quaternion operator can be combined with multiple octonion physical quantities to become a few composite operators. It means that the physical quantities that the quaternion operator can act on are diverse, achieving some octonion physical quantities related to the contributions of different physical quantities (Table 5). For a composite operator, the positions of multiple physical quantities are equal to each other. In other words, the field strength or field potential does not occupy a unique central position, compared to other physical quantities within the composite operators.

The quaternion operator $\lozenge$ can deduce multiple composite operators with various octonion physical quantities. By means of these composite operators, it is able to define some octonion physical quantities that are related to the contributions of different physical quantities, including the octonion field source $\mathbb{S}$ , octonion torque $\mathbb{W}$ , and octonion force $\mathbb{N}$ and others.

In the octonion spaces, the octonion field potential $\mathbb{A}$ can be defined as,
\begin{eqnarray}
\mathbb{A} = i ( \lozenge + k_{xx} \mathbb{X} )^\times \circ \mathbb{X} ~,
\end{eqnarray}
where $k_{xx}$ is one coefficient, to satisfy the needs of dimensional homogeneity.

The octonion field strength $\mathbb{F}$ can be defined as,
\begin{eqnarray}
\mathbb{F} = ( \lozenge + k_{ax} \mathbb{X} + k_{aa} \mathbb{A} ) \circ \mathbb{A} ~,
\end{eqnarray}
where $k_{ax}$ is one coefficient, to satisfy the needs of dimensional homogeneity.

The octonion field source $\mathbb{S}$ can be defined as,
\begin{eqnarray}
\mu \mathbb{S} = - ( \lozenge + k_{fx} \mathbb{X} + k_{fa} \mathbb{A} + k_{ff} \mathbb{F} )^\ast \circ \mathbb{F} ~,
\end{eqnarray}
where $k_{fx}$ is one coefficient, to satisfy the needs of dimensional homogeneity.

The octonion torque $\mathbb{W}$ can be defined as,
\begin{eqnarray}
\mathbb{W} = - v_0 ( \lozenge + k_{lx} \mathbb{X} + k_{la} \mathbb{A} + k_{lf} \mathbb{F} + k_{ll} \mathbb{L} ) \circ  \mathbb{L}  ~,
\end{eqnarray}
where $k_{lx}$ and $k_{ll}$ are two coefficients, to meet the requirements of dimensional homogeneity. $\mathbb{L} = ( \mathbb{R} + k_{rx} \mathbb{X} )^\times \circ \mathbb{P}$ . $\mathbb{P} = \mu \mathbb{S} / \mu_g$ .

The octonion force $\mathbb{N}$ is defined as,
\begin{eqnarray}
\mathbb{N} = - ( \lozenge + k_{wx} \mathbb{X} + k_{wa} \mathbb{A} + k_{wf} \mathbb{F} + k_{wl} \mathbb{L} + k_{ww} \mathbb{W} ) \circ \mathbb{W} ~,
\end{eqnarray}
where $k_{wx}$ and $k_{ww}$ are two coefficients, to satisfy the needs of dimensional homogeneity.

In the octonion spaces, these equations in the above can be expanded further, in case certain conditions are met. The expansion methods of Eqs.(21) to (25) are similar to these of Eqs.(5) to (7). For example, one can expand Eq.(25) into eight equations independent of each other, by replacing the composite operator $( i \mathbb{F} / v_0 + \lozenge )$ with $( \lozenge + k_{wx} \mathbb{X} + k_{wa} \mathbb{A} + k_{wf} \mathbb{F} + k_{wl} \mathbb{L} + k_{ww} \mathbb{W} )$ . In particular, the expansion method of Eq.(25) is similar to that of Eq.(7).

\begin{table}[h]
\centering
\caption{The quaternion operator, octonion field potential, and octonion field strength can be combined together to become a few composite operators, achieving several octonion physical quantities related to the contribution of multiple physical quantities.}
\begin{tabular}{@{}ll@{}}
\hline\noalign{\smallskip}
octonion physical quantity~~~~~~~  &   definition                                                                                               \\
\noalign{\smallskip}\hline\noalign{\smallskip}
octonion field potential           &   $\mathbb{A} = i ( \lozenge + k_{xx} \mathbb{X} )^\times \circ \mathbb{X}$                                \\
octonion field strength            &   $\mathbb{F} = ( \lozenge + k_{ax} \mathbb{X} + k_{aa} \mathbb{A} ) \circ \mathbb{A}$                     \\
octonion field source              &   $\mu \mathbb{S} = - ( \lozenge + k_{fx} \mathbb{X} + k_{fa} \mathbb{A} + k_{ff} \mathbb{F} )^\ast \circ \mathbb{F}$  \\
octonion torque                    &   $\mathbb{W} = - v_0 ( \lozenge + k_{lx} \mathbb{X} + k_{la} \mathbb{A} + k_{lf} \mathbb{F} + k_{ll} \mathbb{L} )
                                            \circ  \mathbb{L} $    \\
octonion force                     &   $\mathbb{N} = - ( \lozenge + k_{wx} \mathbb{X} + k_{wa} \mathbb{A} + k_{wf} \mathbb{F} + k_{wl} \mathbb{L}
                                            + k_{ww} \mathbb{W} ) \circ \mathbb{W}$              \\
\noalign{\smallskip}\hline
\end{tabular}
\end{table}

\section{\label{sec:level1}Discussions and conclusions}

The quaternion space $\mathbb{H}_e$ can be utilized to describe the physical properties of electromagnetic fields, and the quaternion space $\mathbb{H}_g$ may be applied to explore the physical properties of gravitational fields. Two independent quaternion spaces, $\mathbb{H}_g$ and $\mathbb{H}_e$ , can be combined together to become an octonion space $\mathbb{O}$ . So the octonion spaces are able to study simultaneously the physical properties of gravitational and electromagnetic fields.

In the octonion spaces, the octonion field strength and quaternion operator will constitute a composite operator. Making use of the composite operator related with the contributions of octonion field strength, it is capable of deducing the octonion field source, linear momentum, angular momentum, torque and force. In certain special cases, the octonion force may be equal to zero. When the octonion force is zero, eight independent equations can be derived from the equation of octonion force, including the force equilibrium equation, fluid continuity equation, current continuity equation, and precession equilibrium equation and so forth. Similarly, the octonion field potential and quaternion operator may compose a few composite operators, in the octonion spaces. When the octonion field potential plays a significant role, these composite operators can be utilized to deduce the definitions of octonion field source, linear momentum, angular momentum, torque and force.

For diverse combinations of octonion physical quantities and quaternion operator, it is able to achieve a few different composite operators. (a) A composite operator composed of octonion field strength and quaternion operator. (b) Some composite operators posed of octonion field potential and quaternion operator. (c) Several composite operators constituted of octonion field potential, field strength, and quaternion operator.

Each physical quantity, acted on by the quaternion operator $\lozenge$ , is supposed to be equal before the composite operator in the octonion field theory. These physical quantities and quaternion operator can constitute various composite operators. As the physical quantities acted on by the quaternion operator vary, the composite operators can be extended in different ways.

Multiple octonion physical quantities and quaternion operator can also form some composite operators, in the octonion spaces. By means of these composite operators, it is capable of deducing the octonion field source, linear momentum, angular momentum, torque and force, when several octonion physical quantities exert an influence.

It is worth noting that the paper investigates only several special cases of composite operators, composed of multiple octonion physical quantities and quaternion operator, in the octonion spaces. But it has clearly demonstrated that the octonion field strength and field potential are able to exert a significant impact on the gravitational and electromagnetic equations. Some other important octonion physical quantities may also contribute to the electromagnetic and gravitational equations to a certain extent. In the future research, we shall investigate the physical properties of a few special cases of composite operators, related to various octonion physical quantities, in the octonion spaces.

\section*{Acknowledgements}

The author is indebted to the anonymous referees for their valuable comments on the previous manuscripts. This project was supported partially by the National Natural Science Foundation of China under grant number 60677039.


\begin{thebibliography}{99}







\bibitem{flaut}
      Flaut, C., Shpakivskyi, V.,
      ``An Efficient Method for Solving Equations in Generalized Quaternion and Octonion Algebras",
      {\it Advances in Applied Clifford Algebras\/},
      Vol. 25, No. 2, pp. 337--350, 2015.

\bibitem{gogberashvili}
      Gogberashvili, M.,
      ``Octonionic version of Dirac equations",
      {\it International Journal of Modern Physics A\/},
      Vol. 21, No. 17, pp. 3513--3523, 2006.

\bibitem{goldman}
      Goldman, R.,
      ``Understanding quaternions",
      {\it Graphical Models\/},
      Vol. 73, No. 02, pp. 21--49, 2011.

\bibitem{rawat}
      Rawat,~A.~S., Negi,~O.~P.~S.,
      ``Quaternion gravi-electromagnetism",
      {\it International Journal of Theoretical Physics\/},
      Vol.~51, No.~03, pp. 738--745, 2012.

\bibitem{mironov}
      Mironov,~V.~L., Mironov,~S.~V.,
      ``Octonic representation of electromagnetic field equations",
      {\it Journal of Mathematical Physics\/},
      Vol.~50, id. 012901, 2009.

\bibitem{deleo}
      De~Leo,~S., Ducati~G.,
      ``The octonionic eigenvalue problem"
      {\it Journal of Physics A-Mathematical and Theoretical\/},
      Vol.~45, No.~31, id. 315203, 2012.

\bibitem{tanisli1}
      Tanisli, M., Kansu, M. E., Demir, S.,
      ``Reformulation of electromagnetic and gravito electromagnetic equations for Lorentz system with octonion algebra",
      {\it General Relativity and Gravitation\/},
      Vol. 46, No. 05, id. 1739, 2014.

\bibitem{moffat}
      Moffat,~J.~W.,
      ``Higher-dimensional Riemannian geometry and quaternion and octonion spaces",
      {\it Journal of Mathematical Physics\/},
      Vol.~25, No.~02, pp. 347--350, 1984.

\bibitem{morita}
      Morita,~K.,
      ``Quaternions, Lorentz group and the Dirac theory",
      {\it Progress of Theoretical Physics\/},
      Vol.~117, No.~03, pp. 501--532, 2007.

\bibitem{bernevig}
      Bernevig, B. A., Hu, J.-Q., Toumbas, N., and Zhang, S.-C.,
      ``Eight dimensional quantum Hall effect and octonions",
      {\it Physical Review Letters\/},
      Vol. 91, No. 23, id. 236803, 2003.

\bibitem{furui}
      Furui,~S.,
      ``Axial anomaly and the triality symmetry of octonion",
      {\it Few-Body Systems\/},
      Vol.~54, No.~11, pp. 2097--2111, 2013.

\bibitem{chanyal2}
      Chanyal, B. C., Bisht, P. S., Li, T.-J., Negi, O. P. S.,
      ``Octonion quantum chromodynamics",
      {\it International Journal of Theoretical Physics\/},
      Vol. 51, No. 11, pp. 3410--3422, 2012.

\bibitem{furey}
      Furey, C.,
      ``Generations: three prints, in colour",
      {\it Journal of High Energy Physics\/},
      Vol. 2014, No. 10, id. 046, 2014.

\bibitem{majid}
      Majid, S.,
      ``Gauge theory on nonassociative spaces",
      {\it Journal of Mathematical Physics\/},
      Vol. 46, No. 10, id. 103519, 2005

\bibitem{farrill}
      Figueroa-O'Farrill, J. M.,
      ``Gauge theory and the division algebras",
      {\it Journal of Geometry and Physics\/},
      Vol. 32, No. 02, pp. 227--240, 1999

\bibitem{bossard}
      Bossard, G.,
      ``Octonionic black holes",
      {\it Journal of High Energy Physics\/},
      Vol. 2012, No. 05, id. 113, 2012.

\bibitem{tanisli2}
      Demir,~S., Tanisli,~M.,
      ``Hyperbolic octonion formulation of the fluid Maxwell equations",
      {\it Journal-Korean Physical Society\/},
      Vol. 68, No. 05, pp. 616--623, 2016.

\bibitem{kansu}
      Demir,~S., Tanisli,~M., Kansu,~M.~E.,
      ``Generalization of compressible fluid equations in terms of complexified octonions",
      {\it International Journal of Geometric Methods in Modern Physics\/},
      Vol. 20, No. 12, id. 2350211, 2023.

\bibitem{demir}
      Demir,~S., Zeren,~E.,
      ``Multifluid plasma equations in terms of hyperbolic octonions",
      {\it International Journal of Geometric Methods in Modern Physics\/},
      Vol. 15, No. 04, id. 1850053, 2018.


\bibitem{weng3}
      Weng, Z.-H.,
      ``Eight equilibrium and continuity equations within the material media",
      {\it International Journal of Modern Physics A\/},
      Vol. 37, No. 1, id. 2250004, 2022.

\bibitem{weng1}
      Weng, Z.-H.,
      ``Angular momentum and torque described with the complex octonion",
      {\it AIP Advances\/},
      Vol. 4, No. 8, id. 087103, 2014.
      (Erratum, Vol. 5, No. 10, id. 109901 (2015).

\bibitem{weng2}
      Weng, Z.-H.,
      ``Gauge fields and four interactions in the trigintaduonion spaces",
      {\it Mathematical Methods in the Applied Sciences\/},
      Vol. 47, id. MMA10345, 2024(DOI: 10.1002/mma.10345)(in press).

\bibitem{a-b}
      Aharonov, Y., Bohm D.,
      ``Significance of electromagnetic potentials in quantum theory",
      {\it Physical Review\/},
      Vol. 115, No. 3, pp. 485--491, 1959.

\bibitem{tonomura}
      Osakabe, N., Matsuda, T., Kawasaki, T., Endo, J., Tonomura, A., Yano, S., Yamada, H.,
      ``Experimental confirmation of Aharonov-Bohm effect using a toroidal magnetic field confined by a superconductor",
      {\it Physical Review A\/},
      Vol. 34, No. 2, pp. 815--822, 1986.

\bibitem{berry}
      Berry, M. V.,
      ``Quantal phase factors accompanying adiabatic changes",
      {\it Proceedings of the Royal Society of London A\/},
      Vol. 392, No. 1802, pp. 45--57, 1984.

\bibitem{hohensee}
      Hohensee, M. A., Estey, B., Hamilton, P., Zeilinger, A., Muller, H.,
      ``Force-Free Gravitational Redshift: Proposed Gravitational Aharonov-Bohm Experiment",
      {\it Physical Review Letters\/},
      Vol. 108, No. 23, id. 230404, 2012.

\bibitem{overstreet}
      Overstreet, C., Asenbaum, P., Curti, J., Kim, M., Kasevich, M. A.,
      ``Observation of a gravitational Aharonov-Bohm effect",
      {\it Science\/},
      Vol. 375, No. 6577, pp. 226--229, 2022.

\bibitem{weng4}
      Weng, Z.-H.,
      ``Conserved quantities of vectorial magnitudes within the material media",
      {\it International Journal of Modern Physics A\/},
      Vol. 38, No. 3, id. 2350001, 2023.






\end{thebibliography}
\end{document}